\documentstyle[12pt]{article}

\input epsf
\begin{document}
\begin{titlepage}
\today          \hfill
\begin{center}
\hfill    OITS-671 \\

\vskip .05in

{\large \bf Determining the Weak Phase $\gamma$ 
using the decays
$B_d$, $B^+ \rightarrow K \eta \; ( \eta ^{\prime} )$ and
$B_s \rightarrow \pi \eta \; ( \eta ^{\prime} )$
}
\footnote{This work is supported by DOE Grant DE-FG03-96ER40969.}
\vskip .15in
K. Agashe \footnote{email: agashe@oregon.uoregon.edu} and N.G. Deshpande
\footnote{email: desh@oregon.uoregon.edu}
\vskip .1in
{\em
Institute of Theoretical Science \\
University
of Oregon \\
Eugene OR 97403-5203}
\end{center}

\vskip .05in

\begin{abstract}
We suggest two methods (based on flavor
$SU(3)$ symmetry) to determine the CKM angle
$\gamma$ using the decays
$B_d$, $B^+ \rightarrow K \eta \; (\eta ^{\prime})$ and
$B_s \rightarrow \pi \eta \; (\eta ^{\prime})$, respectively.
Rescattering effects
are partly included -- we neglect annihilation
amplitudes, but do not assume any other relation between the
$SU(3)$ invariant amplitudes.
We use the fact that the amplitude (including the Electroweak Penguin
contribution)
for $B_d$, $B^+ \rightarrow
\pi K$ with final state $I \; (\hbox{isospin}) \;
=3/2$ is known as a function of $\gamma$ from the
decay rate $B^+ \rightarrow \pi^0 \pi^+$. 
\end{abstract}

\end{titlepage}

\newpage
\renewcommand{\thepage}{\arabic{page}}
\setcounter{page}{1}

\section{Introduction}
The principal aim of the $B$ physics experimental programs is to
measure the angles (denoted
by
$\alpha$, $\beta$ and $\gamma$) of the triangle 
representing the unitarity relation:
$V^
{\ast}_{tb} V_{td} +
V^{\ast}_{cb} V_{cd} + V^{\ast}_{ub} V_{ud} = 0$, where $V$ is the
Cabibbo-Kobayashi-Maskawa (CKM) matrix.
The idea is 
to overdetermine the angles of this triangle
and thus test the CKM
paradigm of CP violation. 

Methods to determine $\gamma$ ($\equiv$
Arg $(- V^{\ast}_{ub} V_{ud} / V^{\ast}_{cb} V_{cd}))$
including the Electroweak Penguin (EWP) diagram
contribution have been 
suggested \cite{ghlr, dh,gr,nr1,others}.
To (over)determine $\gamma$ to test the CKM theory of CP violation, we 
measure $\gamma$ using different techniques. It is thus
useful to have new methods to determine $\gamma$.
With this motivation, 
in this letter, we give two new methods to determine $\gamma$ using
time {\em integrated} rates for the decays
$B_d$, $B^+ \rightarrow K \eta \; ( \eta ^{\prime} )$ (Method 1)
and $B_s \rightarrow \pi \eta \; ( \eta^{\prime} )$ (Method 2). 
As in all the other methods mentioned above, flavor $SU(3)$ symmetry 
is used in both the methods.

We will write the decay amplitudes in terms of flavor $SU(3)$ invariant
amplitudes \cite{d}. These are denoted by
$C^{T,P}_3$, $C^{T,P}_6$, $C_{15}^{T,P}$, $A_3^{T,P}$ and
$A_{15}^{T,P}$ and correspond to the $5$ linearly independent ways of
forming flavor $SU(3)$ singlets from the initial meson $B_i$, the
two final state mesons belonging to the flavor $SU(3)$ octet
and the effective weak Hamiltonian which
transforms as a $\bar{3} \times 3 \times \bar{3}$. $T$ and $P$ denote
the parts of these amplitudes generated by tree level and penguin 
operators respectively.
These invariant amplitudes include soft final state rescattering effects.
Some of the methods to determine $\gamma$ \cite{nr1}, 
including the ones which use the decays $B^+ \rightarrow K \eta 
\; ( \eta ^{\prime} )$ \cite{dh,gr} and the decay $B_s \rightarrow \pi
\eta \; ( \eta ^{\prime} )$  
\cite{ghlr} neglect rescattering effects. 
In particular,
these methods assume that the decay amplitude $B^+ \rightarrow \pi^+
K^0$ has no weak phase $e^{i \gamma}$ from the tree level operators. In the
language of the  
$SU(3)$ invariant amplitudes, this is equivalent to assuming that the 
annihilation amplitudes \footnote{Annihilation amplitudes
are the ones in which the index $i$ of $B_i$ is contracted directly with
the Hamiltonian.} $A_i$ are suppressed by $f_B/m_B$ 
and 
a combination of the $SU(3)$ invariant amplitudes,
$C_3^T - C^T_6 - C^T_{15}$ is zero (both of 
which are valid in the absence of significant rescattering effects).
Rescattering effects can enhance the annihilation contributions
{\em and} lead to significant $C_3^T - C^T_6 - C^T_{15}$ \cite{rescatt}.
In this letter, we 
neglect annihilation contributions
but {\em do not} assume any relation
between $C^T_3$, $C^T_6$ and $C_{15}^T$ or 
the other $SU(3)$
invariant amplitudes.
Thus, rescattering effcts are partly included.

The decay amplitudes for
$B_d \rightarrow \pi K$ can be written as \cite{d}
\begin{eqnarray}
-{\cal A} (B_d \rightarrow
\pi^-  K^+) &  = & \lambda _u ^{(s)} (C_3^T + C_6^T+
3 C_{15}^T ) + \sum _q \lambda _q ^{(s)} \left(C_{3,q}^P + C_{6,q}^P+
3 C_{15,q}^P \right) \nonumber \\
 & & - \lambda _u ^{(s)} A_{15}^T - \sum _q \lambda _q^{(s)} A_{15,q}^P,
\nonumber \\
\sqrt{2} \; {\cal A} (B_d \rightarrow
\pi^0  K^0) & = & \lambda _u ^{(s)} (C_3^T + C_6^T -
5 C_{15}^T ) + \sum _q \lambda _q ^{(s)} \left(C_{3,q}^P + C_{6,q}^P -
5 C_{15,q}^P \right)
\nonumber \\
 & & - \lambda _u ^{(s)} A_{15}^T - \sum _q \lambda _q^{(s)} A_{15,q}^P.
\label{b0pik}
\end{eqnarray}
Here, $\lambda _q ^{(q^{\prime})} = V^{\ast}_{q b} V_{q q^{\prime}}$
($q = u,c,t$ and $q^{\prime} = d,s$) and $C_q^P$, $A^P_q$ denote the penguin
amplitudes due to quark $q$ running in the loop.
Using the unitarity of the CKM matrix, {\it i.e.,} $\lambda _t ^{(s)}
= - \lambda _u ^{(s)} - \lambda _c ^{(s)}$, we get
\begin{equation}
\lambda _u ^{(s)} C^T_i + \sum_q \lambda _q ^{(s)} C_{i,q}^P =
\lambda _u ^{(s)} \tilde{C}^T_i - \lambda _c ^{(s)} C_i^P,
\label{not}
\end{equation}
where $\tilde{C}^T _{i} =C_{i}^T - C_{i,t}^P + C_{i,u}^P$ and
$C_{i}^P = C_{i,t}^P - C_{i,c}^P$.
A simliar notation is used for $\tilde{A}_i^T$ and $A_i^P$. Henceforth, we
will write the decay amplitudes using this notation.

The decay amplitudes for $B^+ \rightarrow \pi K$ are
\cite{d}
\begin{eqnarray}
{\cal A} (B^+ \rightarrow \pi^+ K^0) & = & 
\lambda _u ^{(s)} \tilde{C}_3^T - \lambda _c ^{(s)} C_{3}^P - 
\lambda _u ^{(s)} \tilde{C}_6^T + \lambda _c ^{(s)} C_6^P 
\nonumber \\
 & & - 
\lambda _u ^{(s)} \tilde{C}_{15}^T + \lambda _c ^{(s)} C_{15}^P
+ 3 \lambda _u ^{(s)} \tilde{A}_{15}^T - 3 \lambda _c ^{(s)} A_{15}^P,
\nonumber \\ 
- \sqrt{2} \; {\cal A} (B^+ \rightarrow \pi^0 K^+) & = & 
\lambda _u ^{(s)} \tilde{C}_3^T - \lambda _c ^{(s)} C_{3}^P - 
\lambda _u ^{(s)} \tilde{C}_6^T + \lambda _c ^{(s)} C_6^P 
\nonumber \\ 
 & & + 7 \lambda _u ^{(s)} \tilde{C}_{15}^T - 7 \lambda _c ^{(s)} C_{15}^P
+ 3 \lambda _u ^{(s)} \tilde{A}_{15}^T - 3 \lambda _c ^{(s)} A_{15}^P.
\nonumber \\
 & & 
\label{b+pik}
\end{eqnarray}
Using Eqns.(\ref{b0pik}) and (\ref{b+pik}), we get an expression for
$A_{3/2}$, the amplitude for $B^+$, $B_d \rightarrow \pi K$ with final state
$I \; (\hbox{isospin}) \; = 3/2$,
\begin{eqnarray}
A_{3/2} & = & {\cal A} (B^+ \rightarrow \pi^+ K^0) + 
\sqrt{2} \; {\cal A} (B^+ \rightarrow \pi^0 K^+) \nonumber \\
 & = & {\cal A} (B_d \rightarrow
\pi^-  K^+) + \sqrt{2} \; {\cal A} (B_d \rightarrow
\pi^0  K^0) \nonumber \\
 & = & - 8 \; \left( \lambda _u ^{(s)} \tilde{C}_{15}^T - \lambda _c ^{(s)}
C_{15}^P \right).
\label{a3/2}
\end{eqnarray}
The decay amplitude for $B^+ \rightarrow \pi^+ \pi^0$ is
\begin{equation}
- \sqrt{2} {\cal A} (B^+ \rightarrow \pi^+ \pi^0) = 
8 \; (\lambda _u ^{(d)} \tilde{C}_{15}^T - \lambda _c ^{(d)} C_{15}^P ).
\end{equation}
The QCD penguin diagram (which is $\Delta I =1/2$)
does not contribute to this decay since this decay has the transition
$\Delta I =3/2$. So, $C^P_{15}$ is the EWP contribution.
Neubert, Rosner \cite{nr2} showed 
that 
\begin{equation}
C_{15,q}^P = C_{15}^T \; \frac{3}{2} \;
\kappa _q,
\end{equation}
where $\kappa _q = \; (c_{9,q} + c_{10,q}) / (c_1 + c_2)$ is the ratio
of Wilson coefficients (WC's) of the
EWP operators (with quark $q$ running in the loop) and the tree
level operators in the effective Hamiltonian
so that 
\begin{equation}
- \sqrt{2} {\cal A} (B^+ \rightarrow \pi^+ \pi^0) = 8 \; 
C_{15}^T \left[ \lambda _u ^{(d)} \left( 1 + \frac{3}{2} \kappa_u
- \frac{3}{2} \kappa_t  \right)  - \lambda _c ^{(d)}
\left( \frac{3}{2} \kappa_t 
- \frac{3}{2} \kappa_c \right) \right].
\end{equation}
The top quark EWP diagram with $Z$ exchange
is enhanced by $m^2_t/ m_Z^2$
and so $\kappa_t \gg \kappa_{u,c}$ giving
\begin{equation}
- \sqrt{2} {\cal A} (B^+ \rightarrow \pi^+ \pi^0) \approx  
8 \; C_{15}^T \lambda _u ^{(d)} \left[ \left(1 - \frac{3}{2} \kappa
\right) - \frac{\lambda _c ^{(d)}}
{\lambda _u ^{(d)}} \frac{3}{2} \kappa \right], 
\end{equation}
where $\kappa = \kappa_t$.
Since $3/2 \; \kappa \sim 2 \%$ and $| \lambda _u ^{(d)}| \sim | 
\lambda _u ^{(d)}|$, we get
\begin{equation}
- \sqrt{2} {\cal A} (B^+ \rightarrow \pi^+ \pi^0) \approx 8 \; C_{15}^T
| \lambda _u ^{(d)} | e^{i \gamma}
\end{equation}
in the Wolfenstein parametrization and setting the strong phase of
$C_{15}^T$ to zero, {\it i.e.,} the EWP contribution 
is $\sim O(2 \%)$ and 
can thus be neglected in the decay amplitude $B^+ \rightarrow \pi^0
\pi^+$. \footnote{This EWP contribution can actually 
be included \cite{others}, but we neglect it for simplicity here.}
Thus, $C^T_{15}$ can be determined directly from the decay rate
$B^+ \rightarrow \pi^+ \pi^0$.

Similarly, the expression for $A_{3/2}$ (Eqn.(\ref{a3/2}))
simplifies to
\begin{equation}
A_{3/2} \approx - 8 \; C_{15}^T \left( | \lambda _u ^{(s)} | e^{i \gamma}
- | \lambda _c ^{(s)} | \frac{3}{2} \kappa \right) 
\end{equation}
so that
\begin{equation}
| A_{3/2} | = 8 \; C^T_{15} | \lambda _u ^{(s)} |
\sqrt{ \left( 1 + \delta _{EW} ^2 - 2 \delta _{EW} \cos \gamma \right) },
\label{a3/2nr}
\end{equation}
where
$\delta _{EW}$ is given by $|\lambda _c ^{(s)}|
/ |\lambda _u ^{(s)}| \; 3/2 \; \kappa
\sim O(1) $,
{\it i.e.,} 
in this case, due to the CKM factors, the EWP contribution 
$\propto \lambda _c ^{(s)}$ {\em is} important.
Thus, knowing $C^T_{15}$ from the decay rate
$B^+ \rightarrow \pi^+ \pi^0$ and $\gamma$, 
we can determine $A_{3/2}$ and conversely $\gamma$
can be determined if the (magnitude) of $A_{3/2}$ is known using some other
method and the decay rate $B^+ \rightarrow \pi^0 \pi^+$ is measured. 
In using this relation, it is crucial that
the parameter $\delta _{EW}$ is calculable. 

In the analysis up to now, annihilation contributions are included.

\section{Method 1}
\label{b+b0}
The decay amplitudes for $B_d$, $B^+$
decays to $\eta_8 = 1/ \sqrt{6} \left(  2 \; s \bar{s} - u \bar{u} - d \bar{d}
\right)$ and $\eta_1 = 1/ \sqrt{3} \left(
s \bar{s} + u \bar{u} + d \bar{d} \right)$ can be written as \cite{d}
\begin{eqnarray}
\sqrt{6} {\cal A} \left( B^+ \rightarrow K^+ \eta _8 \right)
& = & \lambda _u ^{(s)} \tilde{C}^T_{3} - \lambda _c ^{(s)} C^P_{3}
- \lambda _u ^{(s)} \tilde{C}^T_{6} + \lambda _c ^{(s)} C^P_{6} \nonumber \\
 & & - 9 \; \left( \lambda _u ^{(s)} \tilde{C}^T_{15} 
- \lambda _c ^{(s)} C^P_{15} \right)
+ 3 \; \left( 
\lambda _u ^{(s)} \tilde{A}^T_{15} - \lambda _c ^{(s)} A^P_{15} \right),
\nonumber \\
\sqrt{3} {\cal A} \left( B^+ \rightarrow K^+ \eta _1 \right)
& = & 2 \; \left( 
\lambda _u ^{(s)} \tilde{C}^T_{3} - \lambda _c ^{(s)} C^P_{3} \right)
+ \lambda _u ^{(s)} \tilde{C}^T_{6} - \lambda _c ^{(s)} C^P_{6} \nonumber \\
 & & + 3 \; \left( \lambda _u ^{(s)} \tilde{C}^T_{15} 
- \lambda _c ^{(s)} C^P_{15} \right)
+ 6 \; \left( 
\lambda _u ^{(s)} \tilde{A}^T_{15} - \lambda _c ^{(s)} A^P_{15} \right) 
\nonumber \\
 & & 
+ 3 \; \left( 
\lambda _u ^{(s)} \tilde{E}^T_3 - \lambda _c ^{(s)} E^P_3 \right)
+ 3 \; \left( \lambda _u ^{(s)} \tilde{D}^T_{6} - \lambda _c ^{(s)} 
D^P_6 \right) \nonumber \\
 & & + 9 \left( \lambda _u ^{(s)} \tilde{D}^T_{15}- \lambda _c ^{(s)}
D^P_{15} \right),
\label{b+keta81}
\end{eqnarray}
\begin{eqnarray}
\sqrt{6} {\cal A} \left( B_d \rightarrow K^0 \eta _8 \right)
& = & \lambda _u ^{(s)} \tilde{C}^T_{3} - \lambda _c ^{(s)} C^P_{3}
+ \lambda _u ^{(s)} \tilde{C}^T_{6} - \lambda _c ^{(s)} C^P_{6} \nonumber \\
 & & - 5 \; \left( \lambda _u ^{(s)} \tilde{C}^T_{15}
- \lambda _c ^{(s)} C^P_{15} \right)
- \left(
\lambda _u ^{(s)} \tilde{A}^T_{15} - \lambda _c ^{(s)} A^P_{15} \right),
\nonumber \\
\sqrt{3} {\cal A} \left( B_d \rightarrow K^0 \eta _1 \right)
& = & 2 \; \left(
\lambda _u ^{(s)} \tilde{C}^T_{3} - \lambda _c ^{(s)} C^P_{3} \right)
- \lambda _u ^{(s)} \tilde{C}^T_{6} + \lambda _c ^{(s)} C^P_{6} \nonumber \\
 & & - \left( \lambda _u ^{(s)} \tilde{C}^T_{15}
- \lambda _c ^{(s)} C^P_{15} \right)
- \; 2 \; \left(
\lambda _u ^{(s)} \tilde{A}^T_{15} - \lambda _c ^{(s)} A^P_{15} \right)
\nonumber \\
 & & + 3 \; 
\left( \lambda _u ^{(s)} \tilde{E}^T_3 - \lambda _c ^{(s)} E^P_3 \right)
- 3 \; \left( \lambda _u ^{(s)} \tilde{D}^T_{6} - \lambda _c ^{(s)}
D^P_6 \right) \nonumber \\
 & & - 3 \; \left( \lambda _u ^{(s)} \tilde{D}^T_{15}- \lambda _c ^{(s)}
D^P_{15} \right),
\label{b0keta81}
\end{eqnarray}
where $E_3$, $D_6$ and $D_{15}$ are the amplitudes which contribute only to
$B$ meson decays to a final state involving $\eta _1$
\cite{d}. \footnote{The notation $\tilde{D}^T$, $D_P$ is similar to 
$\tilde{C}^T$, $C_P$ (Eqn.(\ref{not})).} $D_6$ and $D_{15}$
are annihilation amplitudes.
We assume that the mass eigenstates $\eta$ and $\eta ^{\prime}$ 
are given by the
the canonical mixing \footnote{Both the methods
can be easily modified in the case of a general mixing angle, provided
the mixing angle is known.}:
\begin{eqnarray}
\eta & = & \frac{2 \sqrt{2}}{3} \; \eta_8 - \frac{1}{3} \; \eta_1 \nonumber \\
 & = & \frac{1}{\sqrt{3}} \left( s \bar{s} - u \bar{u} - d \bar{d} \right),
\nonumber \\
\eta ^{\prime} & = & \frac{1}{3} \; \eta_8 + \frac{2 \sqrt{2}}{3} \; \eta_1
\nonumber \\
 & = & \frac{1}{\sqrt{6}} \left( 2 \; s \bar{s} + u \bar{u} + d \bar{d} 
\right).
\label{etamix}
\end{eqnarray}
This mixing is consistent
with the present data \cite{etaphen}.
Then,
the decay amplitudes for $B^+$, $B_d$ decays to $\eta$ and $\eta^{\prime}$
are
\begin{eqnarray}
- \sqrt{3} \; {\cal A} \left( B^+ \rightarrow K^+ \eta \right)
& = & 
\lambda _u ^{(s)} \tilde{C}^T_{6} - \lambda _c ^{(s)} C^P_{6} 
+ 7 \; \left( \lambda _u ^{(s)} \tilde{C}^T_{15}
- \lambda _c ^{(s)} C^P_{15} \right) \nonumber \\ & & 
+ \left( \lambda _u ^{(s)} \tilde{E}^T_3 - \lambda _c ^{(s)} E^P_3 \right)
+ \; \left( \lambda _u ^{(s)} \tilde{D}^T_{6} - \lambda _c ^{(s)}
D^P_6 \right) \nonumber \\
 & & 
+ 3 \; \left( \lambda _u ^{(s)} \tilde{D}^T_{15}- \lambda _c ^{(s)}
D^P_{15} \right), \nonumber \\
\sqrt{6} \; {\cal A} \left( B^+ \rightarrow K^+ \eta ^{\prime} \right)
& = & 3 \; \left(
\lambda _u ^{(s)} \tilde{C}^T_{3} - \lambda _c ^{(s)} C^P_{3} \right)
+ \lambda _u ^{(s)} \tilde{C}^T_{6} - \lambda _c ^{(s)} C^P_{6} \nonumber \\
 & & + \left( \lambda _u ^{(s)} \tilde{C}^T_{15}
- \lambda _c ^{(s)} C^P_{15} \right)
+ 9 \; \left(
\lambda _u ^{(s)} \tilde{A}^T_{15} - \lambda _c ^{(s)} A^P_{15} \right)
\nonumber \\
 & & + 4 \;
\left( \lambda _u ^{(s)} \tilde{E}^T_3 - \lambda _c ^{(s)} E^P_3 \right)
+ 4 \; \left( \lambda _u ^{(s)} \tilde{D}^T_{6} - \lambda _c ^{(s)}
D^P_6 \right) \nonumber \\
 & & + 12 \; \left( \lambda _u ^{(s)} \tilde{D}^T_{15}- \lambda _c ^{(s)}
D^P_{15} \right),
\label{b+keta}
\end{eqnarray}
\begin{eqnarray}
- \sqrt{3} \; {\cal A} \left( B_d \rightarrow K^0 \eta  \right)
& = & 
- \lambda _u ^{(s)} \tilde{C}^T_{6} + \lambda _c ^{(s)} C^P_{6} 
+3 \;  \left( \lambda _u ^{(s)} \tilde{C}^T_{15}
- \lambda _c ^{(s)} C^P_{15} \right)
\nonumber \\ 
& & 
+ \left( \lambda _u ^{(s)} \tilde{E}^T_3 - \lambda _c ^{(s)} E^P_3 \right)
- \; \left( \lambda _u ^{(s)} \tilde{D}^T_{6} - \lambda _c ^{(s)}
D^P_6 \right) \nonumber \\
 & &  
- \; \left( \lambda _u ^{(s)} \tilde{D}^T_{15}- \lambda _c ^{(s)}
D^P_{15} \right), \nonumber \\
\sqrt{6} \; {\cal A} \left( B_d \rightarrow K^0 \eta ^{\prime} \right)
& = & 
3 \; \left( 
\lambda _u ^{(s)} \tilde{C}^T_{3} - \lambda _c ^{(s)} C^P_{3} \right)
- \lambda _u ^{(s)} \tilde{C}^T_{6} + \lambda _c ^{(s)} C^P_{6} \nonumber \\
 & & - 3 \; \left( \lambda _u ^{(s)} \tilde{C}^T_{15}
- \lambda _c ^{(s)} C^P_{15} \right)
- 3 \; \left(
\lambda _u ^{(s)} \tilde{A}^T_{15} - \lambda _c ^{(s)} A^P_{15} \right)
\nonumber \\
 & & + 4 \;
\left( \lambda _u ^{(s)} \tilde{E}^T_3 - \lambda _c ^{(s)} E^P_3 \right)
- 4 \; \left( \lambda _u ^{(s)} \tilde{D}^T_{6} - \lambda _c ^{(s)}
D^P_6 \right) \nonumber \\
 & & - 4 \; \left( \lambda _u ^{(s)} \tilde{D}^T_{15}- \lambda _c ^{(s)}
D^P_{15} \right).
\label{b0keta}
\end{eqnarray}
From Eqns.(\ref{b+pik}),
(\ref{b+keta81})
and (\ref{etamix}), we get the relation \cite{dh}
\begin{eqnarray}
\sqrt{6} {\cal A} \left( B^+ \rightarrow K^+ \eta _8\right) & = & 
\frac{4}{3} \sqrt{3} {\cal A} \left( B^+ \rightarrow K^+ \eta \right) +
\frac{1}{3} \sqrt{6} {\cal A} \left( B^+ \rightarrow K^+ \eta ^{\prime}
\right) \nonumber \\
& = & 2 \; {\cal A} \left( B^+ \rightarrow \pi^+ K^0 \right) 
+ \sqrt{2} {\cal A} \left( B^+ \rightarrow \pi^0 K^+ \right).
\nonumber \\
 & & 
\label{b+pikketa}
\end{eqnarray}

\begin{figure}
\centerline{\epsfxsize=1\textwidth \epsfbox{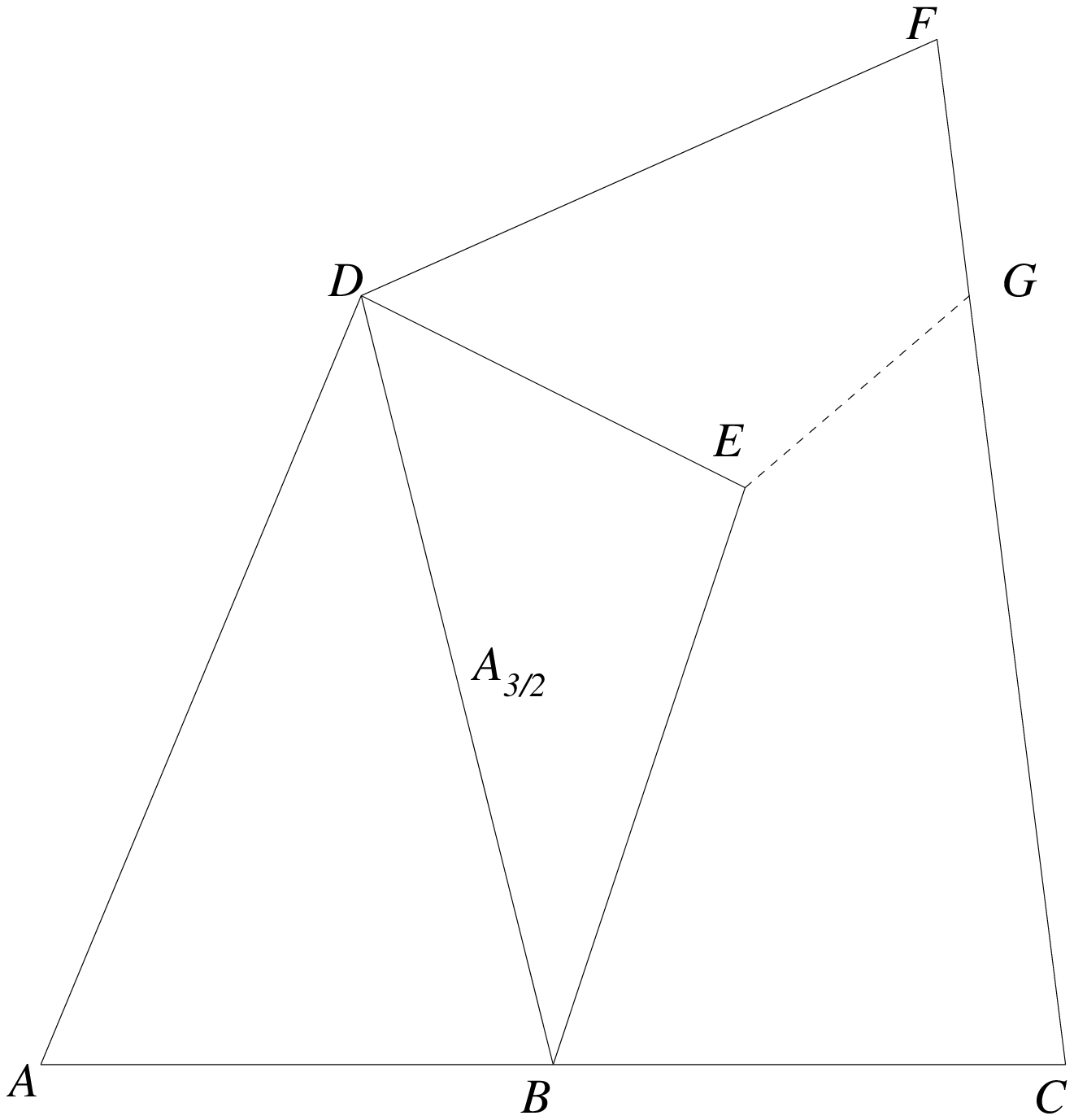}}
\caption{The polygon construction for Method 1. $AB = BC = 
|{\cal A}
\left( B^+ \rightarrow \pi^+ K^0 \right)|$ , $AD = 
\protect\sqrt{2} \; |{\cal A}
\left( B^+ \rightarrow \pi^0 K^+\right)| $, $DF = 
\protect\frac{1}{3} \protect\sqrt{6} \; | {\cal A}
\left( B^+ \rightarrow K^+ \eta ^{\prime}\right)| $, 
$GC = 3 \; FG = \protect\sqrt{3} \; |{\cal A}
\left( B^+ \rightarrow K^+ \eta \right)|$, $DE =
\protect\sqrt{2} \; |{\cal A}
\left( B_d \rightarrow \pi^0 K^0 \right)| $,
$BE = |{\cal A}
\left( B_d \rightarrow \pi^- K^+ \right)| $. Given $\gamma$ and
the $B^+ \rightarrow \pi^+ \pi^0$ decay rate we know
$DB = | A_{3/2} |$ from Eqn.(\protect\ref{a3/2nr}). The prediction for
$\protect\sqrt{3} \; | {\cal A}
\left( B_d \rightarrow K^0 \eta \right) | $ is $EG$.}
\protect\label{figbketa}
\end{figure}

As mentioned earlier, the magnitude of $A_{3/2}$ is known as a function of
$\gamma$ in terms of the $B^+ \rightarrow \pi^0 \pi^+$ decay rate
(Eqn.(\ref{a3/2nr})).
This, for a given $\gamma$, we can construct the two triangles 
formed by $B_d \rightarrow \pi K$, $A_{3/2}$
and $B^+ \rightarrow \pi K$, $A_{3/2}$ (corresponding
to Eqns.(\ref{a3/2}) and $\Delta$'s
$DEB$ and $ADB$ respectively
in Fig.\ref{figbketa}) and the quadrangle formed
by $B^+ \rightarrow \pi K$, $K^+ \eta$, $K^+ \eta ^{\prime}$
corresponding to
Eqn.(\ref{b+pikketa}) ($ADFC$ of Fig.\ref{figbketa}). \footnote
{There are two discrete ambiguities in the construction of
Fig.\ref{figbketa}: $\Delta$'s $ADB$ and $DEB$ can be on the same side
of the common base $DB$ and similarly in the quadrangle $ADFC$ the 
vertices $A$ and $F$ can be on the same side of the diagonal $DC$.}
Thus, we know the phases (in the convention where
the phase of $C^T_{15}$ is zero) of the decay 
amplitudes $B^+ \rightarrow \pi K$, $\eta K^+$, $K^+ \eta ^{\prime}$
and $B_d \rightarrow
\pi K$ as a function of $\gamma$ form this construction
(the magnitudes are, of course, known
from the measurement of the decay rates).

We have included the annihilation contributions up to now.

If we {\em neglect} the annihilation amplitudes, $A_{15}$, $D_6$ and $D_{15}$,
using Eqns.(\ref{b0pik}), (\ref{b+keta}) and (\ref{b0keta}),
we get the relations:
\begin{eqnarray}
- \; \sqrt{3} {\cal A} \left( B_d \rightarrow K^0 \eta \right) 
& = & - \; \sqrt{2} {\cal A} \left( B_d \rightarrow \pi^0 K^0 \right) 
+ \frac{1}{3} \sqrt{6} \left( B^+ \rightarrow K^+ \eta ^{\prime}
\right) \nonumber \\
 & & + \frac{1}{3} \sqrt{3} {\cal A}
\left( B^+ \rightarrow K^+ \eta \right),
\label{b+b0etaKrel1}
\end{eqnarray}
\begin{eqnarray}
\sqrt{6} {\cal A} \left( B_d \rightarrow K^0 \eta ^{\prime} \right) 
& = & - \; \sqrt{2} {\cal A} \left( B_d \rightarrow \pi^0 K^0 \right) 
+ \frac{4}{3} \sqrt{6} \left( B^+ \rightarrow K^+ \eta ^{\prime}
\right) \nonumber \\
 & & + \frac{4}{3} \sqrt{3} {\cal A}
\left( B^+ \rightarrow K^+ \eta \right).
\label{b+b0etaKrel2}
\end{eqnarray}
Thus, we can predict  
the decay amplitudes $B_d \rightarrow K^0 \eta$,
$K^0 \eta^{\prime}$ as a function of $\gamma$ since, as mentioned above,
we know the magnitudes and phases 
(the latter from Fig.\ref{figbketa})
of all the other decay amplitudes
in Eqns.(\ref{b+b0etaKrel1}) and (\ref{b+b0etaKrel2}).  
In fact, the decay amplitude
$B_d \rightarrow K^0 \eta$ is shown as $EG$ in Fig.\ref{figbketa}.
Once the decay rate $B_d \rightarrow K^0 \eta$ or $K^0 \eta^{\prime}$
is measured, $\gamma$ can be determined.
Thus, $\gamma$ can be determined (up to a four-fold discrete
ambiguity) by the
measurement of the decay rates for $8$ modes -- $B^+ 
\rightarrow \pi^0 \pi^+$, $\pi K$, $K^+ \eta$, $K^+ \eta ^{\prime}$, $B_d
\rightarrow \pi K$, $K^0 \eta$ (or $K^0 \eta ^{\prime}$).

\section{Method 2}
\label{b+b0bs}
This method is based on the method of Gronau {\it et al.}
\cite{ghlr}. In \cite{ghlr}, the annihilation amplitudes are neglected
{\em and} the relation $C^T_3 - C^T_6 - C_{15}^T = 0$ is assumed. In other
words, rescattering effects are neglected so that the amplitude
for the decay $B^+ \rightarrow K^0 \pi^+$ has no weak phase $e^{i \gamma}$.
We neglect the annihilation amplitudes, but
do not assume any relation between the $C$'s. Thus, we include
partly the rescattering effects.
The decay amplitudes for $B_s \rightarrow \pi^0 \eta _8$, $\eta _1$
are \cite{d}
\begin{eqnarray}
\sqrt{3} {\cal A} (B_s \rightarrow \pi^0 \eta _8) & = &
2 \left( \lambda _u ^{(s)} \tilde{C}_{6}^T - \lambda _c ^{(s)} C^P_{6}
\right) - 
4 \left( \lambda _u ^{(s)} \tilde{C}_{15}^T - \lambda _c ^{(s)} C^P_{15}
\right) \nonumber \\
 & & + 4 \left( \lambda _u ^{(s)} \tilde{A}_{15}^T - \lambda _c ^{(s)} A^P_{15}
\right),
\label{bseta8}
\end{eqnarray}
\begin{eqnarray}
\sqrt{6} {\cal A} (B_s \rightarrow \pi^0 \eta_1) & = &
2 \left( \lambda _u ^{(s)} \tilde{C}_{6}^T - \lambda _c ^{(s)} C^P_{6}
\right) - 4 \left( \lambda _u ^{(s)} \tilde{C}_{15}^T - \lambda _c ^{(s)} C^P_{15}
\right) \nonumber \\
 & & 
-8  \left( \lambda _u ^{(s)} \tilde{A}_{15}^T - \lambda _c ^{(s)} A^P_{15}
\right) + 6 \lambda _u ^{(s)} \tilde{D}^T_{6} - 6 \lambda _c ^{(s)} D^P_{6}
\nonumber \\ 
 & & - 12 \lambda _u ^{(s)} \tilde{D}^T_{15} + 12
\lambda _c ^{(s)} D^P_{15}.
\label{bseta1}
\end{eqnarray}
With the canonical mixing (Eqn.\ref{etamix})), we get
\begin{eqnarray}
\sqrt{6} {\cal A} (B_s \rightarrow \pi^0 \eta) & = &
2 \left( \lambda _u ^{(s)} \tilde{C}_{6}^T - \lambda _c ^{(s)} C^P_{6}
\right) 
- 4 \left( \lambda _u ^{(s)} \tilde{C}_{15}^T - \lambda _c ^{(s)} C^P_{15}
\right) \nonumber \\
 & & 
+ 8 \left( \lambda _u ^{(s)} \tilde{A}_{15}^T - \lambda _c ^{(s)} A^P_{15}
\right)
- 2 \left( \lambda _u ^{(s)} \tilde{D}^T_{6} -
\lambda _c ^{(s)} D^P_{6} \right)
\nonumber \\
 & & + 4 \left(
\lambda _u ^{(s)} \tilde{D}^T_{15} - \lambda _c ^{(s)} D^P_{15}
\right),
\label{bseta}
\end{eqnarray}
\begin{eqnarray}
{\cal A} (B_s \rightarrow \pi^0 \eta ^{\prime}) & = &
\sqrt{2} (B_s \rightarrow \pi^0 \eta) - 2 \sqrt{3} \left[
2 \left( \lambda _u ^{(s)} \tilde{A}_{15}^T - \lambda _c ^{(s)} A^P_{15}
\right) \right. \nonumber \\
 & & - \left. \lambda _u ^{(s)} \tilde{D}^T_{6} +
\lambda _c ^{(s)} D^P_{6} + 2 \left(
\lambda _u ^{(s)} \tilde{D}^T_{15} - \lambda _c ^{(s)} D^P_{15}
\right) \right].
\label{bseta2}
\end{eqnarray}
Neglecting the annihilation ampltudes, $A_{15}$, $D_6$ and $D_{15}$, 
from Eqns.(\ref{b0pik}), (\ref{b+pik}), 
(\ref{bseta}) and (\ref{bseta2}) we get the relations \cite{ghlr}:
\begin{eqnarray}
\sqrt{6} {\cal A} \left( B_s \rightarrow \pi^0 \eta \right) & \approx &
- {\cal A} \left( B^+ \rightarrow K^0 \pi^+ \right) + \sqrt{2}
{\cal A} \left( B_d \rightarrow K^0 \pi^0 \right) \nonumber \\
 & = & \sqrt{2} {\cal A} \left( B^+ \rightarrow K^+ \pi^0 \right) -
{\cal A} \left( B_d \rightarrow K^+ \pi^- \right),
\label{bsb0b+rel}
\end{eqnarray}
\begin{equation}
\sqrt{6} {\cal A} \left( B_s \rightarrow \pi^0 \eta \right) \approx
\sqrt{3} {\cal A} \left( B_s \rightarrow \pi^0 \eta ^{\prime} \right).
\label{etaetaprime}
\end{equation}

\begin{figure}
\centerline{\epsfxsize=1.25\textwidth \epsfbox{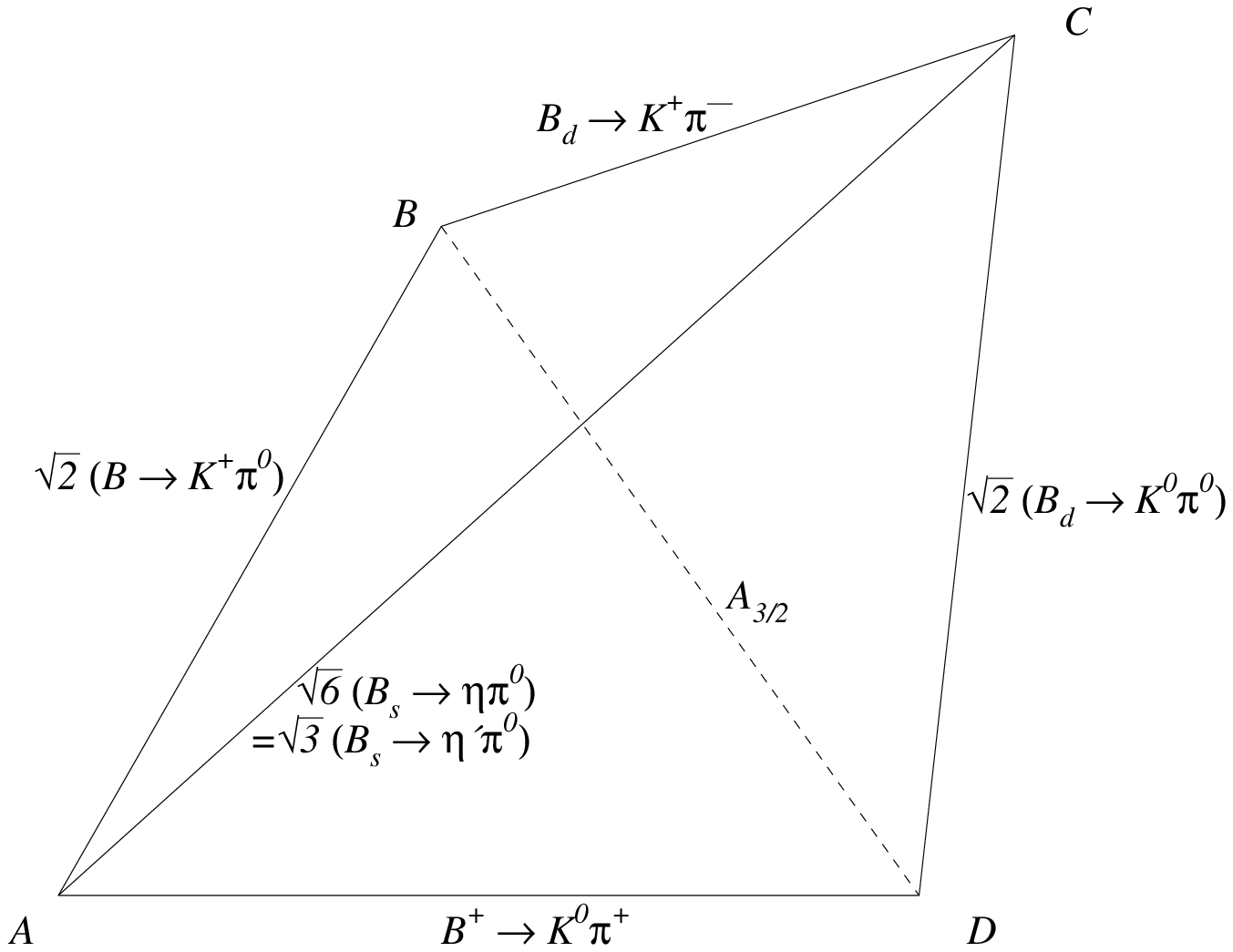}}
\caption{
The polygon construction for Method 2. Knowing $| A_{3/2} |$ from 
this Figure, we can determine $\gamma$ using Eqn.(\protect\ref{a3/2nr})
if the decay rate $B^+ \rightarrow \pi^0 \pi^+$ is also measured.}
\protect\label{figbspieta}
\end{figure}

From Eqns.(\ref{bsb0b+rel}) and (\ref{etaetaprime})
we see that 
the decay amplitudes $B_d \rightarrow \pi K$, $B^+ \rightarrow \pi K$ form 
the sides 
and $B_s \rightarrow \pi \eta$ (or $\eta^{\prime}$) the diagonal of
a quadrangle shown in Fig.\ref{figbspieta}.
Thus, the measurement of these
$5$ decay rates fixes this quadrangle \footnote{
There is a discrete ambiguity in this construction
since $\Delta$'s $ABC$ and $ADC$ can be on the same
side of the common base $AC$ in Fig.\ref{figbspieta}.} 
of which the other diagonal is 
$A_{3/2}$ (see Eqns.(\ref{a3/2}) and Fig.\ref{figbspieta}) \cite{ghlr}. 
Knowing the magnitude of $A_{3/2}$ from Fig.\ref{figbspieta}
and the decay rate $B^+ \rightarrow \pi^0 \pi^+$, we can determine
$\gamma$ from Eqn.(\ref{a3/2nr}). 
Thus, $\gamma$ can be determined (up to a two-fold
discrete ambiguity) by the
measurement of the decay rates for $6$ modes -- $B^+
\rightarrow \pi^0 \pi^+$, $\pi K$, $B_d
\rightarrow \pi K$, and $B_s \rightarrow
\pi^0 \eta$ (or $\pi^0 \eta ^{\prime}$).
In the method of
\cite{ghlr}, measurement of the rates for 
the CP-conjugate processes of all the above
modes is also 
required to determine $\gamma$.

\section{Discussions}
We comment on the accessibility of the various decay modes used in the two 
methods. The $B_d$ and $B^+$ decay modes should be accessible at the $e^+ e^-$
machines whereas the $B_s \rightarrow \pi \eta \; (\eta ^{\prime})$
decay mode will only be accessible at a hadron machine. Since
the QCD penguin does not contribute to this $B_s$ decay,
the decay rate is expected to be small. The measurements of the
decay rates to CP 
eigenstate final states: $B_s \rightarrow \pi \eta \; (\eta ^{\prime})$,
$B_d \rightarrow \pi^0 K^0$ and $B_d \rightarrow K^0 \eta \; (\eta ^{\prime})$
require external tagging.

As mentioned earlier, we have used flavor $SU(3)$ symmetry in both the
methods. 
In the factorization approximation,
$SU(3)$ breaking in the tree level amplitudes can be incorporated by factors
of $f_K / f_{\pi}$ (see, for example,
\cite{ghlr}). For example, 
$C^T_{15} \; (\Delta S = 1) = f_K / f_{\pi} \times
C^T_{15} \; (\Delta S =0)$.
However, since some of the strong penguin operators are
$(V-A) \times (V+A)$,
in the penguin amplitudes, the $SU(3)$ breaking
effects are difficult to estimate, but the breaking will still be less than
$\sim O(30 \%)$. In method 2, we use the decay mode
$B_s \rightarrow \pi
\eta \; (\eta^{\prime})$ which does not have the QCD penguin
contribution, but does have the 
EWP contribution. The EWP operators ${\cal O}_{7,8}$ have very small
WC's whereas the EWP operators with significant WC's, ${\cal O}_{9,10}$,
are Fierz-equivalent to the tree level operators ${\cal O}_{1,2}$
\cite{nr2}. So
in the factorization approximation,
the corrections due to $SU(3)$ breaking in relating the penguin
amplitudes for $B_s \rightarrow \pi
\eta \; (\eta^{\prime})$ to the ones for $B_d$, $B^+ \rightarrow 
\pi K$ are 
given by factors of $f_{\eta,\eta^{\prime}}
/ f_K$.  

We have also assumed that the $SU(3)$ breaking in the strong phases
is small. A possible justification is that at the energies
of the final state particles $\sim m_b/2$, the phase shifts are
not expected to be sensitive to the $SU(3)$ breaking given by,
say, $m_K - m_{\pi}$ (which is much smaller than
the final state momenta). However, it is hard to
quantify this effect.

Both the methods can be used with CP-conjugates of all the decay modes
as well.
We have neglected annihilation amplitudes: $A_{15}$, $D_6$ and $D_{15}$.
The validity of this assumption can be
checked by comparing the decay rates $B_s \rightarrow \pi \eta$
and $B_s \rightarrow \pi \eta ^{\prime}$ -- these two decay amplitudes
differ only in the annihilation contribution (see Eqn.(\ref{bseta2})).
In the absence of significant annihilation contribution, the decay rate for
$B_s \rightarrow \pi \eta ^{\prime}$ should be twice that for 
$B_s \rightarrow \pi \eta$.

In summary, we have discussed two new methods (based on flavor
$SU(3)$ symmetry) to determine
the weak phase $\gamma$ using
the decays $B_d$, $B^+ \rightarrow K \eta$ ($\eta ^{\prime}$)
and $B_s \rightarrow \pi \eta$ $(\eta ^{\prime})$, respectively. These methods
partly take into account rescattering effects.

\end{document}